\documentclass[prl,twocolumn,showpacs]{revtex4}
\usepackage{graphicx}
\usepackage{bm}
\usepackage{times}
\usepackage{epsfig}

\def\be{\begin{equation}}
\def\ee{\end{equation}}
\def\bea{\begin{eqnarray}}
\def\eea{\end{eqnarray}}
\def\bse{\begin{subequations}}
\def\ese{\end{subequations}}

\def\be{\begin{eqnarray}}
\def\ee{\end{eqnarray}}

\begin{document}

\title{Effects of quasiparticle ambipolarity on the Nernst effect in underdoped
cuprate superconductors}
\author{Sumanta Tewari$^{1}$}
\author{Chuanwei Zhang$^2$}
\affiliation{$^{1}$Department of Physics and Astronomy, Clemson University, Clemson, SC
29634\\
$^2$Department of Physics and Astronomy, Washington State University,
Pullmann, WA 99164}

\begin{abstract}
We consider the Nernst effect in the normal state of the underdoped cuprate superconductors, where
recent quantum oscillation experiments have indicated the existence of Fermi
surface pockets and quasiparticle ambipolarity in the excitation spectrum.
Using an ambipolar $d$-density wave model for the pseudogap, we predict a
peak in the Nernst coefficient with decreasing temperature below the
pseudogap temperature. The existence of the peak and its sign,
which we predict to be the same as that due to mobile vortices in a superconductor, result from the dominance of the electron
pocket at low temperatures, as has been observed
in recent Hall experiments.
\end{abstract}

\pacs{74.72.-h, 72.15.Jf, 72.10.Bg}
\maketitle

Physics of the normal state of the cuprate high-temperature superconductors
in the intermediate range of hole doping, called underdoping, is an
important unresolved problem \cite{Norman}.
In this doping range,
 the system evinces a gap in the spectrum  of unidentified origin
(pseudogap) below a temperature scale $T^{*} > T_c$. An understanding of this pseudogap phase is widely
believed to hold the key to the high transition temperature in the cuprates \cite{Norman}.
Encouragingly, recent quantum oscillation experiments~\cite%
{Doiron-Leyraud:2007,LeBoeuf:2007,Bangura:2008,Jaudet:2008,Yelland:2008,Sebastian:2008} have found evidence of a remnant Fermi surface, consisting of small pockets, even in the enigmatic pseudogap phase. This has ushered in the encouraging prospect of describing
this phase in terms of well defined low energy quasiparticles of a state with a broken symmetry and a reconstructed
Fermi surface \cite{LeBoeuf:2007,Chakravarty:2008b,Millis:2007,Podolsky:2008}. In
this paper, we will derive the implications of these important new ingredients in
the cuprate physics on a widely discussed transport coefficient -- Nernst coefficient -- using a well-known model of the pseudogap
-- the ambipolar $d$-density wave (DDW) model \cite{Nayak,Chakravarty:2001} -- which is consistent with the findings of the quantum oscillation experiments.
 It will be shown that the reconstructed Fermi surface and its low energy quasiparticles
may help resolve another long-standing puzzle in the cuprate physics, that involving an enhanced Nernst signal at temperatures much above
$T_c$ \cite{Wang1,Lee2,Wang2,Olivier}.
Enhanced Nernst effect in the cuprates has been discussed before in terms of possible vortex motion in the
pseudogap phase \cite{Podolsky} and the superconducting fluctuation scenario \cite{Serbyn}. While these approaches are expected to produce a
large peak in the Nernst coefficient close to $T_c$, we show here that the quasiparticle ambipolarity associated with
a competing ordered state can extend the onset temperature of a sizable Nernst signal to temperatures much higher
than $T_c$.

The quantum oscillation experiments indicate that the Fermi surface in the underdoped cuprates is made
of small pockets, giving
rise to both hole and electron-like charge carriers (quasiparticle ambipolarity) in the excitation spectrum.
Remarkably, the primary oscillation frequency
 is consistent with the existence of
an electron-like pocket \cite{Taillefer} in the excitation spectrum.  This is despite the fact that the
systems are actually moderately hole doped. A resolution of this puzzle may lie in the
relative mobilities  of the hole and the electron-like carriers: the
oscillation frequency of the former may be strongly suppressed due to lower
mobilities at low temperature ($T$) \cite{LeBoeuf:2007}.  An even more
convincing evidence that this may indeed be the case comes from the sign of
the low-$T$, normal state  Hall coefficient $R_H$ \cite{LeBoeuf:2007,Taillefer}.  
In recent experiments \cite{LeBoeuf:2007,Taillefer} on YBCO, $R_H$ has been
observed to become negative  for $T < T_0 < T^{*}$.
This points to the dominance of the electron pocket at $T < T_0$ ($T_0$: compensation temperature) over the
closed hole-like parts of the reconstructed Fermi surface.
We will derive the implications of the ambipolar flow of the oppositely-charged
quasiparticles, and the low-$T$ dominance of the electron pocket, on
Nernst coefficient using the ambipolar DDW model of the pseudogap as an illustrative example.

\begin{figure}[t]
\includegraphics[width=0.65\linewidth]{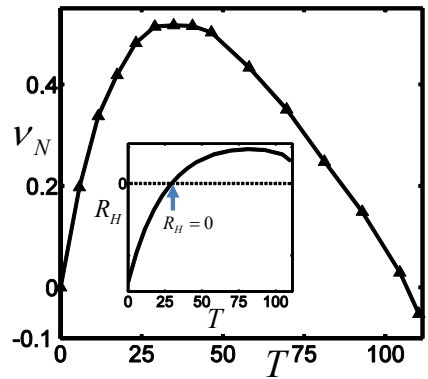}
\caption{Nernst coefficient, $\protect\nu _{N}$, in arbitrary units
versus $T$ in the ambipolar DDW state.
The inset shows the Hall coefficient, $R_{H}$, calculated for the same set of
ambipolar DDW parameters. Here, $T^{\ast }=110$ K, $A_{e}=58$ K, $A_{h}=$ 253 K, $t=0.3$ eV, $%
t^{\prime }=0.3t$, $B_{e}=B_{h}=1$, $W_{0}=0.1$ eV, $\protect\mu =-0.26$ eV, which corresponds to $x \sim 0.1$.}
\label{fig:Nernst}
\end{figure}

The ordered DDW model, as others ~\cite%
{Chakravarty:2008b,Millis:2007,Podolsky:2008}, has enjoyed some
success in explaining the quantum oscillation experiments in the pseudogap
regime.
Based on this illustrative model of quasiparticle ambipolarity, the principal result of this paper is that the Nernst
coefficient shows a pronounced peak with decreasing $T$ in the pseudogap
phase, as shown in Fig.~(\ref{fig:Nernst}).
We also find that the sign of the low-$T$ Nernst coefficient
is the same as that due to vortices in a superconductor \cite{Behnia}.
This sign is opposite to that above $%
T^*$, and should be construed as a consequence of the low-$T$
dominance of the electron pocket in the quasiparticle spectrum.
The Nernst
coefficient for the DDW state was studied earlier
\cite{Oganesyan}, but a pronounced low temperature peak with positive sign as discussed here was not found.

The commensurate DDW state \cite{Nayak} is described by an order parameter
which is a particle-hole singlet in spin space,
\begin{equation}
\left\langle \hat c_{\bm k+\bm Q,\alpha}^{\dagger} \hat c_{\bm k,\beta}
\right\rangle \propto iW_{\bm k}\,\delta_{\alpha\beta}, \; W_{\bm k}=\frac{
W_0}{2}(\cos k_x-\cos k_y).  \label{Order-Parameter}
\end{equation}
Here $\hat c^{\dagger}$ and $\hat c$ are the electron creation and
annihilation operators on the square lattice of the copper atoms, $\bm
k=(k_x,k_y)$ is the two-dimensional momentum, $\bm Q=(\pi,\pi)$ is the wave
vector of the density wave, and $\alpha$ and $\beta$ are the spin indices.
In Eq.~(\ref{Order-Parameter}), $iW_{\bm k}$ is the DDW order parameter with
the $id_{x^2-y^2}$ symmetry in the momentum space. For $\bm
Q=(\pi,\pi)$, it is purely imaginary \cite{Nayak} and gives rise to
spontaneous currents along the bonds of the square lattice.

The Hartree-Fock Hamiltonian describing the mean-field DDW state is given by,
\begin{eqnarray}
&\hat{H}=\sum_{{\bm k}\in \mathrm{RBZ}}\left(
\begin{array}{cc}
\varepsilon _{\bm k}-\mu & iW_{\bm k} \\
-iW_{\bm k} & \varepsilon _{\bm k+\bm Q}-\mu%
\end{array}
\right) ,&  \label{Hamiltonian} \\
&\varepsilon _{\bm k}=-2t(\cos k_{x}+\cos k_{y})+4t^{\prime }\cos k_{x}\cos
k_{y},&
\end{eqnarray}
where $\varepsilon _{\bm k}$ is the band dispersion of the electrons, and $%
\mu$ is the chemical potential. The Hamiltonian density in Eq.~(\ref%
{Hamiltonian}) operates on the two-component spinor $\hat{\Psi}_{\bm k}=(%
\hat{c}_{\bm k}, \hat{c}_{\bm k+\bm Q})$ defined on the reduced Brillouin
zone (RBZ) described by $k_x \pm k_y = \pm \pi$, and can be expanded over
the Pauli matrices $\hat{ \bm \tau }$ and the unity matrix $\hat{I}$, $\hat{%
H_{\bm k}}=w_{0}(\bm k)\hat{I}+\bm w(\bm k)\cdot \hat{\bm\tau },\quad w_{0}=
\frac{\varepsilon _{\bm k}+\varepsilon _{\bm k+\bm Q}}{2}-\mu$, $%
w_{1}=0,\quad w_{2}=-W_{\bm k},\quad w_{3}=\frac{ \varepsilon _{\bm %
k}-\varepsilon _{\bm k+\bm Q}}{2}$. 
The spectrum of the Hamiltonian consists of two branches with the
eigenenergies $E_{\pm }(\bm k)=w_{0}(\bm k) \pm w(\bm k)$, where $w(\bm k)=|%
\bm w(\bm k)|$. For a generic set of parameters, corresponding to a non-zero
hole doping in the underdoped regime of the cuprates, the reconstructed
Fermi surface consists of two hole pockets near the $(\pi /2,\pm \pi /2)$
points and one electron pocket near the $(\pi ,0)$ point in the reduced
Brillouin zone. The existence of both hole and electron-like excitations in
the quasiparticle spectrum makes this state an ambipolar state.

In some recent Nernst experiments \cite{Wang1,Lee2,Wang2}, a temperature gradient, $-%
\mathbf{\nabla }T$, is applied along the $\hat{x}$ direction ($T$ \emph{decreases} along $\hat{x}$).
For such a temperature gradient, and
with a magnetic field $\mathbf{B}$ along the $\hat{z}$ direction, the vortices of a superconductor close to $T_c$ produce
a transverse electric field in the positive $\hat{y}$ direction. With the sign convention
that this signal is positive,
the Nernst coefficient can be written as,
\begin{equation}  \label{eq:1Q_xy}
\nu_{N} = \frac{E_y}{(- \nabla T)_x B} = \frac{\alpha_{xy} \sigma_{xx}
-\alpha_{xx} \sigma_{xy}}{\sigma_{xx}^2+\sigma_{xy}^2},
\end{equation}
where $\sigma _{ij}$ and $\alpha _{ij}$ are the
electric and the thermoelectric conductivity tensors, respectively.

We calculate the off-diagonal element of the conductivity tensor, $\sigma
_{xy}$, by using the solution of the semi-classical Boltzmann
equation \cite{Trugman}:
\begin{eqnarray}
\sigma_{xy}(\mu) &=& e^{3}B\tau_e^{2}\int\frac{d^{2}k}{(2\pi)^{2}}
\Big[\frac{\partial E_{+}(k)}{\partial k_{x}}\frac{\partial E_{+}(k)}{\partial
k_{y}}
\frac{\partial^{2}E_{+}(k)}{\partial k_{x} \partial k_{y}}\nonumber\\ &-&
\left(\frac{\partial E_{+}(k)}{\partial
k_{x}}\right)^{2}\frac{\partial^{2}E_{+}(k)}{\partial k_{y}^{2}}\Big]
(-\frac{\partial f(E_{+}(k)-\mu)}{\partial E_{+}})\nonumber\\
&+& (E_{+} \rightarrow E_{-}; \tau_e \rightarrow \tau_h).
\label{eq:xy}
\end{eqnarray}
Here, the momentum integrals are over the reduced Brillouin zone. In the DDW
band-structure, the electron pocket near $(\pi ,0)$ is associated with the
upper band, $E_{+}({\bm k})$. The first integral in Eq.~(\ref{eq:xy}),
therefore, embodies the contribution to $\sigma _{xy}$ due to the
electron-like quasiparticles. The electron transport
 scattering time $\tau _{e}$ is
taken as independent of the location on the electron Fermi line. The
second integral in Eq.~(\ref{eq:xy}) with hole scattering time $\tau _{h}$ calculates the contribution to $\sigma _{xy}$
from the hole pockets. In general, there is no obvious reason to expect $\tau
_{e}=\tau _{h}$.
For a consistent interpretation of the Hall effect experiments \cite{LeBoeuf:2007}, it has been
recently argued that the scattering times, which
are directly proportional to the career mobilities, may in fact be
different, and, at least at low $T$, $\tau _{e}>\tau _{h}$.
With the above definition of the
parameters, the diagonal element of the conductivity tensor is given by \cite{Trugman},
\begin{eqnarray}
\sigma_{xx}(\mu) &=& e^{2}\tau_e\int\frac{d^{2}k}{(2\pi)^{2}}
\left(\frac{\partial E_{+}(k)}{\partial k_{x}}\right)^{2}(-\frac{\partial f(E_{+}(k)-\mu)}{\partial E_{+}})\nonumber\\
  &+& (E_{+} \rightarrow E_{-}; \tau_e \rightarrow \tau_h).
\label{eq:xx}
\end{eqnarray}

From the solution of the Boltzmann equation at low $T$, the thermoelectric
tensor $\alpha _{ij}$ is related to the conductivity tensor $\sigma _{ij}$
by the Mott relation \cite{Marder}: 
$\alpha _{ij}=-\frac{\pi ^{2}}{3}\,\frac{k_{B}^{2}T}{e}\,\frac{\partial
\sigma _{ij}}{\partial \mu }$. 
Here $e>0$ is the absolute magnitude of the charge of an electron. Using Eq.~(\ref%
{eq:1Q_xy}), the formula for the Nernst coefficient reduces to,
\begin{equation}
\nu _{N}=-\frac{\pi ^{2}}{3}\,\frac{k_{B}^{2}T}{eB}\,\frac{\partial \Theta
_{H}}{\partial \mu }  \label{eq:1nernst1}
\end{equation}%
Here, $\Theta _{H}=\tan^{-1}(\frac{\sigma _{xy}}{\sigma _{xx}})$.
Using Eqs.~(\ref{eq:xy},\ref{eq:xx},\ref{eq:1nernst1}), and reasonable assumptions given below about $\tau _{e}$ and $\tau _{h}$, we can calculate $\nu_{N}$ as a
function of $T$ in the ambipolar DDW state.

Recently, the normal state Hall coefficient, $R_H = \frac{\sigma_{xy}}{%
(\sigma_{xx})^2}$, has been measured as a function of $T$ in three different
samples of underdoped YBCO \cite{LeBoeuf:2007}.
In all three samples, $R_H$ is positive above a temperature $T_0 < T^{*}$,
which is consistent with the systems being moderately hole doped. Plots of $%
R_H$ as a function of $T$ reveal a change of sign from a positive $R_H$ for $%
T > T_0$ to a negative $R_H$ for $T < T_0$.
By considering the possibility of the coexistence of electron and hole pockets in the normal state Fermi surface,
it was shown before \cite{Kopnin} that the flux flow mechanism in the mixed state of a superconductor ($H << H_{c2}, T < T_c$)
can, in principle, produce a sign reversal of the Hall coefficient. Here, however, we do not focus on the flux flow
mechanism, since the experiments \cite{LeBoeuf:2007} are carried out at much higher magnetic fields ($\sim 50$ T), and,
therefore, the drop in $R_H$ below $T^*$ ($> T_c$) and the subsequent change in sign are argued \cite{LeBoeuf:2007} to be unambiguously a
property of the normal state.
The drop of $R_H$ below $T^{*}$
can be explained naturally if the ordered state in question below $T^{*}$ is
inherently ambipolar, and the mobilities of the oppositely charged
quasiparticles are assumed to be unequal and changing with $T$.
For $\tau _{e}=\tau _{h}$, in
which case the formula for $R_{H}$ is independent of the scattering time, and for a
generic set of parameters ($t,t^{\prime },W_{0}$) consistent with the quantum
oscillation experiments in YBCO \cite{Chakravarty:2008b,Xun}, the size and
curvature of the hole pockets are sufficiently bigger than those of the
electron pocket. This implies that, for $\tau_{e}\sim\tau_{h}$, the sign of the overall $R_{H}$ is positive \cite{Tewari}.
We have checked that reasonable modifications of the band structure
parameters cannot change this conclusion. Therefore, within Boltzmann
theory,
if at high temperatures ($T>T_{0}$) $\tau _{e}\sim \tau _{h}$, the Hall
coefficient is positive. On the other hand, if $\tau _{e}> \tau _{h}$ for %
$T<T_{0}$, $R_{H}$ can become negative at low $T$. Note that a higher
mobility of the electron-like quasiparticles at low $T$ is also consistent
with the frequency observed in the quantum oscillation experiments \cite%
{LeBoeuf:2007,Chakravarty:2008b}. An independent, microscopic, justification of the higher lifetime of the electron-like quasiparticles
at low $T$ and in the presence of a magnetic field follows by considering the scattering of both types of quasiparticles by vortices at low temperatures \cite{Stephen}. Because of the
difference of the effective masses between the DDW quasiparticles near the antinodal and the nodal regions in the Brillouin zone, the electron lifetime due to vortex scattering
can be significantly higher than the hole lifetime at low $T$ in the presence of a magnetic field \cite{Xun,Pallab}.

On the above grounds, we choose the minimal $T$-dependence of the scattering
times: $\tau_{e}^{-1} = A_e + B_e T $ and $\tau_h^{-1} = A_h + B_h T$.
With these choices, we have calculated $R_H$ for the ambipolar DDW state as
a function of $T$ with an assumed mean field $T$-dependence of the order
parameter: $W_0(T)=W_0\sqrt{{1-\frac{T}{T^{*}}}}$ ($T^{*}\sim 110$ K). We
choose the values of the temperature independent parameters $A_e, B_e, A_h,
B_h$ by requiring that $\tau_e(T_0)$ is sufficiently larger than $\tau_h(T_0)
$ for $R_H(T)$ to be zero at $T=T_0 \sim 30$ K and negative below this
temperature. We have checked that our conclusions below for the qualitative
behavior of $\nu$ with $T$ are robust to any reasonable variation of the $T$%
-dependence of $\tau_e, \tau_h$ and $W_0(T)$, as long as it supports the dominance of the electron-like
quasiparticles at low $T$. Our results for $R_H$, with the above
illustrative choice of the parameters, are plotted with $T$ in the inset of
Fig.~(\ref{fig:Nernst}).

We now use Eqs.~(\ref{eq:xy},\ref{eq:xx},\ref{eq:1nernst1}) to calculate $%
\nu _{N}$ as a function of $T$ for a specific value of hole doping $x\sim
10\%$. We use the value of $x$ and the same set of parameters $t,t^{\prime
},W_{0}(T=0)$ to calculate the value of $\mu $, which determines the size
and curvature of the hole and electron pockets. Using a mean field $T$%
-dependence of $W_{0}(T)$ and the phenomenological form of $\tau _{e}(T)$
and $\tau _{h}(T)$ above, we plot in Fig.~(\ref{fig:Nernst}) the calculated $%
\nu _{N}$ in the ambipolar DDW state as a function of $T$.
Two important features of the
qualitative behavior of $\nu_{N}$ with $T$ are clearly visible in Fig.~(\ref{fig:Nernst}%
): First, $\nu _{N}$ has a pronounced peak at about $T\sim 35$ K, which is
close, but not equal to, the compensation temperature $T_{0}=30$ K.
The second important result, which is a
consequence of the dominance of the electron pocket at low $T$, is that $\nu
_{N}$, at least near its peak, is of the same sign (positive) as the vortex signal near $T_c$. (The sign of $\nu_N$ is
negative according to the older, textbook convention \cite{Nolas}.) Both
for $T\sim T^{\ast }$ and $T\sim 0$, $\nu _{N}$ approaches zero, the former
due to nearly perfect balance between the two types of quasiparticles (which
can also make $\nu_N$ slightly negative, depending on the relative magnitudes of $%
\tau _{e}(T^{\ast }),\tau _{h}(T^{\ast })$), and the latter due to the
linearity of $\nu _{N}$ with $T$ at low $T$ (Eq.~(\ref{eq:1nernst1})), which
is a remnant effect of the low-$T$ Mott relation \cite{Marder}. After restoring the $\hbar$ and $c$,
we estimate the peak value of $\nu_N$ in the pseudogap phase to be about $100$ nV/KT, which is
of the same order of magnitude as those observed in the experiments \cite{Wang1,Lee2,Wang2,Olivier}.

Because we have
chosen two different scattering times for the holes and the electrons, it
would be inconsistent to compare the value of $\nu _{N}$ in the ambipolar
DDW state with that for $T>T^{\ast }$. However, we have checked that the
quantity $\frac{\partial \Theta _{H}}{\partial \mu }$ in Eq.~(\ref{eq:1nernst1})
is positive above $T^{\ast }$.
Therefore, as $T$ is decreased, somewhere near $T^{\ast }$ the Nernst coefficient must change
sign from negative to positive, and show a peak as $T$ is decreased further.
The electron (hole) pocket gives a positive (negative) contribution to $\nu_N$ according to the sign convention used here.
Since the electron pocket becomes dominant at low $T$ due to the increase in mobility, $\nu_N$ is positive and rapidly increasing.
However, at low enough $T$, it must approach zero, as shown in Eq.~(\ref{eq:1nernst1}),
which produces a pronounced low temperature peak.
 We have checked that these conclusions are valid
irrespective of the specific choice of the $T$-dependence of the scattering
times, only as long as the mobility of the electron-like quasiparticles is
sufficiently bigger than that of the hole-like quasiparticles to produce a negative $R_H$ at low $T$.

\begin{figure}[t]
\includegraphics[width=0.65\linewidth]{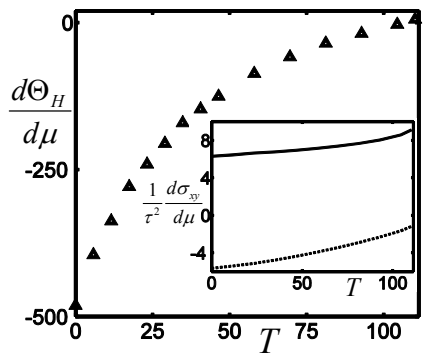}
\caption{ Plot of the quantity $\frac{\partial \Theta _{H}}{%
\partial \protect\mu }$ with $T$ for the ambipolar DDW state. $\protect\nu _{N}$ in Fig.~\ref{fig:Nernst}
is derived from these values by multiplying with $-T$ in eV (Eq.~(%
\protect\ref{eq:1nernst1})). The inset
shows the $T$-dependence of $\frac{\partial \protect\sigma _{xy}}{\partial
\protect\mu }$, which determines the $T$-dependence of $\frac{\partial
\Theta _{H}}{\partial \protect\mu }$, separately for the hole (solid) and
the electron (dotted) pockets.
}
\label{fig:trend}
\end{figure}

The origin of the sign of $\nu _{N}$ can be understood by plotting
the quantity $\frac{\partial \Theta _{H}}{\partial \mu }$, which is
proportional to $\nu _{N}$ (see Eq.~(\ref{eq:1nernst1})), as a function of $T
$. This is shown in Fig.~(\ref{fig:trend}), where this quantity is negative
in the ambipolar DDW state. Writing this quantity as $\frac{\partial \Theta
_{H}}{\partial \mu }=\frac{\frac{\partial \sigma _{xy}}{\partial \mu }\sigma
_{xx}-\frac{\partial \sigma _{xx}}{\partial \mu }\sigma _{xy}}{\sigma
_{xx}^{2}+\sigma _{xy}^{2}}$, we note that the second term in the numerator,
which is proportional to $R_{H}$, but turns out to be small, vanishes at $%
T_{0}$. The $T$-dependence of $\nu _{N}$ is mostly determined by the $T$%
-dependence of $\frac{\partial \sigma _{xy}}{\partial \mu }$, whose values
from the hole and the electron-like pockets are given in the inset of Fig.~(%
\ref{fig:trend}). Since $\sigma _{xy}$ depends sensitively on the curvature
of the Fermi line in 2D, the values of $\frac{\partial \sigma _{xy}}{%
\partial \mu }$ from the individual pockets depend on how the corresponding
curvatures change with small changes in the chemical potential. Since with
an increase in $\mu $ the size and curvature of the electron pocket
increases, the absolute magnitude of $\sigma _{xy}$ from the electron pocket
goes up as well. However, since the sign of $\sigma _{xy}$ is itself
negative from the electron pocket, $\frac{\partial \sigma _{xy}}{\partial
\mu }$ is actually negative for the electron pocket as shown in Fig.~(\ref%
{fig:trend}). By similar reasoning we find that, with an increase in $\mu$,
the curvature of the hole pockets increases where the Fermi velocity is the
largest, increasing the value of $\sigma_{xy}$. Therefore, $\frac{\partial \sigma _{xy}}{\partial \mu }$ is
positive for the hole pockets. Since, with decreasing $T$,
 $\tau _{e}$ becomes larger in comparison to $\tau _{h}$ as indicated by the Hall
  experiments \cite{LeBoeuf:2007}, the effects of the electron-like quasiparticles
become dominant at low $T$, producing a rapid drop in
 $\frac{\partial \Theta _{H}}{\partial \mu }$ with $T$.
 This results in the rapid rise and positive sign of $\nu _{N}(T)$ with $T$. However,
since at lower temperatures $\nu _{N}(T)$ is eventually linear in $T$ (Eq.~(\ref{eq:1nernst1})), and, therefore, approaches zero,
the Nernst coefficient in Fig.~(\ref{fig:Nernst}) has a peak, while $%
\frac{\partial \Theta _{H}}{\partial \mu }$ in Fig.~\ref{fig:trend} does
not.

In conclusion,
 we show that a density-wave mediated Fermi surface reconstruction and quasiparticle ambipolarity
in the underdoped regime of the cuprates
imply a pronounced peak of the Nernst coefficient with decreasing temperature in the pseudogap phase.
Within the ambipolar DDW model of the pseudogap, we find that the sign of the peak Nernst effect is
the same as that expected from the vortices of a superconductor near $T_c$.

We thank S. Chakravarty, L. Taillefer, P. Goswami, and A. Hackl for enlightening discussions. Tewari is supported by DOE/EPSCoR Grant \# DE-FG02-04ER-46139 and  Clemson University start up funds. Zhang is supported
by WSU start-up funds.

\end{document}